\documentstyle[12pt]{article}
\begin{document}

\title{Generation of any superposition of Dicke state of excitons in coupled quantum dots }
\author{XuBo Zou, K. Pahlke and W. Mathis  \\
\\Institute TET, University of Hannover,\\
Appelstr. 9A, 30167 Hannover, Germany }
\date{}

\maketitle

\begin{abstract}
{\normalsize We present a scheme to generate arbitrary
superposition of the Dicke states of excitons in optically driven
quantum dots. This proposal is based on a sequence of laser
pulses, which are tuned appropriately to control transitions on
Dicke state. It is shown that N laser pulses are needed to
generate arbitrary superposition of the Dicke states of N quantum
dots. } PACS number(s): 03.67.-a, 03.67.Hk.
\end{abstract}
Quantum entanglement has been intensely studied, due to its
poential applications in quantum communication and information
processing\cite{ch0} such as quantum teleportation\cite{ch1},
superdense coding\cite{ch2}, quantum key distribution\cite{ch3}
and telecloning\cite{ch4}. One other area where entangled quantum
state may have a significant impact is that of the improvement of
frequency standard\cite{ch5,ch6}. Key to improvement of frequency
standard beyond the short noise limit is the establishment of a
entangled state of a collection of N two level atoms. Initial
theoretical investigation examined the use of spin squeezed
state\cite{ch5,ch6}, which has been extensively studied\cite{ch7}
in recent years. It has also been shown that, in the absence of
decoherence, GHZ state of N the may be used in high precision
spectroscopy to measure the transition frequency to an uncertainty
of $N^{-1}$\cite{ch8, ch9}. When the decoherence is present, the
use of the maximally entangled state does not provide higher
resolution as compared to using independent particle and the best
sensitivity is achieved when the particles are initially prepared
in highly symmetric but only partially entangled states\cite{ch9},
which is a superposition of the Dicke states. A number of
experimental protocols for generation of GHZ state of N qubits and
spin squeezed state have been proposed, including the Cavity
QED\cite{ch10}, trapped ion system\cite{ch11} and Bose-Einstein
condensates\cite{ch7}. In experiment, GHZ state of three or four
qubits has been observed\cite{ch12}. Recently, it has been shown
that there exist two inequivalent classes of three qubit entangled
state under local operation assisted by classical communication,
namely GHZ state $|GHZ>=\frac{1}{\sqrt{2}}(|111>+|000>)$ and W
state $|GHZ>=\frac{1}{\sqrt{3}}(|100>+|010>+|001>)$\cite{ch13}. A
scheme has also been proposed for generating W entangled state of
three or four qubits\cite{ch14}. In this paper, we present a
scheme to generate arbitrary superposition of the Dicke states of
N atoms, which includes GHZ and W state of N qubits, spin squeezed
state and entangled atomic state proposed in Ref\cite{ch9}. Our
scheme is based on optical driven quantum dots. Recent advance in
semiconductor nanostructure fabrication and measurement suggest
that optically generated electron-hole pairs (exciton) in
semiconductor quantum dot represent ideal candidate for achieving
coherence wave function control on the nanometer and femtosecond
scales and implementing a large scale quantum
computation\cite{ch15}. Recently, optically driven quantum dot has
been used to prepare maximally entangled Bell and GHZ
state\cite{ch16} and implement quantum teleportation\cite{ch17}.
In this paper, we present a scheme to generate arbitrary
superposition of the Dicke states of excitons in optically driven
quantum dots. This proposal is based on a sequence of laser
pulses, which are tuned appropriately to control transitions on
Dicke state. It is shown that only one laser pulses are needed to
generate W states of N quantum dots. The schematic requirement are
realizable in current experiment employing ultrafast optical
spectroscopy of quantum dots.\\
In order to describe our scheme, we consider a system of N
identical and equispaced quantum dots containing no net charge
that are radiated by long-wavelength classical light. In the frame
of the rotating wave approximation, the formation of single
excitons within the individual quantum dots and their interdot
hopping are described by the Hamiltonian\cite{ch16,ch17}
$$
H=\Delta_{\omega}J_z+ge^{i\varphi}J_++ge^{-i\varphi}J_-+W(J^2-J_z^2)
\eqno{(1)}
$$
where
$$
J_+=\sum_{p=1}^Nc_p^{\dagger}h_p^{\dagger},~~J_-=\sum_{p=1}^Nc_ph_p
~~J_z=\frac{1}{2}\sum_{p=1}^N(c_p^{\dagger}c_p-h_ph_p^{\dagger})
\eqno{(2)}
$$
Here, $c_p^{\dagger}$ and $c_p$($h_p^{\dagger}$ and $h_p$) are the
electron (hole) creation and annihilation operator in the pth
quantum dot. $\Delta_{\omega}=\epsilon-\omega$. $\epsilon$
represent the band gap, W is the interdot interaction parameter
(Forster process). $\omega$, $\varphi$ and g are frequency, phase
and amplitude of laser field, respectively. The quasispin J
operators satisfy the usual commutation relation $[J_z,
J_{\pm}]=\pm J_{\pm}$, $[J_+, J_{-}]=2J_{z}$ and $[J^2, J_z]=[J^2,
J_{+}]=[J^2, J_{-}]=0$. From a practical point of view, parameter
$\Delta_{\omega}$, $\varphi$ and g are adjustable in the
experiment to give control over the system of the quantum dots. If
the initial state of the system is symmetric with respect to
exchange of quantum dots, We can only consider the symmetric state
and a basis is formed by the Dicke state $|J,M>$, here
$J=\frac{N}{2}, M=-J, -J+1, \cdots, J$. $|J,M>$ is the eigenstate
of the operator $J_z$ with eigenvalue $M$.  Since $J^2$ commute
with $J_z$ and$J_{\pm}$ $J^2$ is conservation constant, we may not
consider the $J^2$ terms in the following. Since the
$\Delta_{\omega}$ is adjustable, in Ref\cite{ch16,ch17}, author
consider resonance condition ($\Delta_{\omega}=0$) and proposed a
scheme to prepare maximally entangled Bell and GHZ state. Here we
choose $\Delta_{\omega}$ to satisfy the condition
$\Delta_{\omega}=-W(2J-2i-1)$, $i=0,1,\cdots,2J-1$. Thus
Hamiltonian of the system can be written in the form
$$
H=-W(J_z+J-i)(J_z+J-i-1)+g_ie^{i\varphi_i}J_++ge^{-i\varphi_i}J_{
-}\eqno{(3)}
$$
Here we have neglected a constant term.  Using the Dicke state
$|J,M>$, $M=-J, -J+1, \cdots, J$ the Hamiltonian(3) take the form
$$
H=-W\sum_{m=-J}^J(m+J-i)(m+J-i-1)|J,m><J,m|
$$
$$
+g_ie^{i\varphi_i}\sum_{m=-J}^{J-1}\sqrt{J(J+1)-m(m+1)}|J,m+1><J,m|
$$
$$
+g_ie^{-i\varphi_i}\sum_{m=-J+1}^{J}\sqrt{J(J+1)-m(m-1)}|J,m-1><J,m|
\eqno{(4)}
$$
It is noticed that the coefficients of the Dicke state $|J,-J+i>$
and $|J,-J+i+1>$ are equal to zero in the first line of above
equation. We now consider the condition $W>>Jg$, which has bee
assumed in Ref\cite{ch16,ch17}. In this case, we apply the
rotating wave approximation and discard the rapidly oscillating
term in the Hamiltonian (4) and obtain the effective interaction
$$
H_i=\Omega_i(e^{i\varphi_i}|J,-J+i+1><J,-J+i|+
e^{-i\varphi_i}|J,-J+i><J,-J+i+1|) \eqno{(5)}
$$
Here $\Omega_i=g\sqrt{J(J+1)-(J-i)(J-i-1)}$. \\
In order to
generate any superposition of Dicke state of excitons
$$
\Psi=\sum_{m=-J}^JC_m|J,m> \eqno{(6)}
$$
we consider the situation in which the quantum dots is prepared in
the zero excitons
$$
\Psi_{initial}=|J,-J> \eqno{(7)}
$$
In the following we show, that each term of equation(1) can be
generated by one laser pulse. Fist, we derive the quantum dots
with laser frequency $\Delta_{\omega}=-W(2J-1)$, after an
interaction time $\tau_0$, the system evolves into
$$
\Psi_{1}=\cos(\Omega_0\tau_0)|J,-J>+i\sin(\Omega_0\tau_0)e^{\varphi_0}|J,-J+1>
\eqno{(8)}
$$
we chhose the amplitude (or interaction time $\tau_0$) of the
laser field in such a way that the following conditional are
fulfilled
$$
\cos(\Omega_0\tau_0)=C_{-J} \eqno{(9)}
$$
we obtain
$$
\Psi_{1}=C_{-J}|J,-J>+ie^{\varphi_0}\sqrt{1-C_{-J}^2}|J,-J+1>
\eqno{(10)}
$$
without loss of generality, we have assumed that $C_{-J}$ is a
real number. We then tune the laser frequency to satisfy
$\Delta_{\omega}=-W(2J-3)$. After an interaction time $\tau_1$,
the system is evolved into
$$
\Psi_{2}=C_{-J}|J,-J>+ie^{\varphi_0}\sqrt{1-C_{-J}^2}(\cos(\Omega_1\tau_1)|J,-J+1>+i\sin(\Omega_1\tau_1)e^{\varphi_1}|J,-J+2>)
\eqno{(11)}
$$
we adjust the amplitude (or interaction time $\tau_1$) of the
second laser field and phase of the laser field of the first laser
field to satisfy
$$
ie^{\varphi_0}\sqrt{1-C_{-J}^2}\cos(\Omega_1\tau_1)=C_{-J+1}
\eqno{(12)}
$$
the state become
$$
\Psi_{2}=C_{-J}|J,-J>+C_{-J+1}|J,-J+1>+i^2e^{i\varphi_0+i\varphi_1}\sqrt{1-C_{-J}^2-|C_{-J+1}|^2}|J,-J+2>
\eqno{(13)}
$$
If this procedure is done for the $m-$th time, the quantum state
of the system is
$$
\Psi_{m}=\sum_{l=-J}^{-J+m-1}C_l|J,l>+i^m\exp(i\sum_{l=0}^{m-1}\varphi_l)\sqrt{1-\sum_{l=-J}^{-J+m-1}|C_1|^2}|J,-J+m>
\eqno{(14)}
$$
We now consider the $(m+1)-$th operation by choosing
$\Delta_{\omega}=-W(2J-2m-1)$. After interaction $\tau_m$, the
quantum state becomes
$$
\Psi_{m+1}=\sum_{l=-J}^{-J+m-1}C_l|J,l>+i^m\exp(i\sum_{l=0}^{m-1}\varphi_l)\sqrt{1-\sum_{l=-J}^{-J+m-1}|C_1|^2}(
\cos(\Omega_m\tau_m)|J,-J+m>
$$
$$
+i\sin(\Omega_m\tau_m)e^{\varphi_m}|J,-J+m+1>) \eqno{(15)}
$$
We choose the amplitude (or interaction time $t_{m}$) of the
(m+1)-th laser pulse and phase of m-th laser field to satisfy
$$
i^m\exp(i\sum_{l=0}^{m-1}\varphi_l)\sqrt{1-\sum_{l=-J}^{-J+m-1}|C_1|^2}
\cos(\Omega_m\tau_m)=C_{-J+m} \eqno{(16)}
$$
After the procedure is performed for $N$ times the system's state
definitely becomes state(6).\\
We present a scheme to generate arbitrary superposition of the
Dicke states of excitons in optically driven quantum dots. This
proposal is based on a sequence of laser pulses, which are tuned
appropriately to control transitions on Dicke state. It is shown
that N laser pulses are needed to generate arbitrary superposition
of the Dicke states of N quantum dots. This scheme of quantum
state generation requires not more than N laser pulses, which is
the smallest possible number. Thus, in respect of short laser
pulse sequences it can be considered as the optimal solution of
this problem of the quantum state generation. it is noticed that
generalized W state of N qubit can be written in the form of Dick
state $|J,-J+1>$, here $J=\frac{N}{2}$, so that it is needed only
one laser pulse to generate W state of N qubits. In this scheme,
we require that $W$ is much larger than amplitude of laser pulse.
This condition is exactly parallel to that in Ref\cite{ch16,ch17},
which can be realized in present experiment. Finally, it is worth
pointing out that the same (or similar) Hamiltonian as in Eq(1).
also arise in other context, i.e. in a two coupled Bose-Einstein
condensate and in the ultrasmall Josephson Junction, which has
been used to create macroscopic quantum superposition and realize
quantum logic gate operation. Thus the present work may be viewed
in a wider context.

\end{document}